\documentclass[aps,superscriptaddress,nofootinbib,preprint,showpacs]{revtex4}
\usepackage{hyperref}
\usepackage{epsfig,rotating}
\usepackage{amsmath,amssymb}
\usepackage{dsfont}

\numberwithin{equation}{section}

\newcommand{\vp}{\vec{p}}

\newcommand{\wu}{\widetilde{U}}
\newcommand{\wh}{\widetilde{H}}
\newcommand{\mf}{\mathcal{F}}
\newcommand{\md}{\mathcal{D}}
\newcommand{\be}{\begin{equation}}
\newcommand{\ee}{\end{equation}}
\newcommand{\bea}{\begin{eqnarray}}
\newcommand{\eea}{\end{eqnarray}}

\begin{document}

\title{Unparticle-Higgs Mixing:\\ MSW Resonances, See-saw Mechanism and Spinodal Instabilities}
\author{D. Boyanovsky}
\email{boyan@pitt.edu} \affiliation{Department of Physics and
Astronomy, University of Pittsburgh, Pittsburgh, PA 15260}
\author{R. Holman} \email{rh4a@andrew.cmu.edu}
\affiliation{Department of Physics, Carnegie Mellon University,
Pittsburgh, PA 15213}
\author{Jimmy A. Hutasoit}\email{jhutasoi@andrew.cmu.edu}  \affiliation{Department of Physics, Carnegie Mellon University, Pittsburgh, PA 15213}

\date{\today}

\begin{abstract}
Motivated by slow roll inflationary cosmology we study  a scalar
unparticle weakly coupled to  a Higgs field in the broken symmetry
phase. The mixing between the unparticle and the Higgs field results
in a see-saw type matrix and the mixing angles feature a
Mikheyev-Smirnov-Wolfenstein (MSW) effect as a consequence of the
unparticle field being non-canonical. We find \emph{two} (MSW)
resonances for small and large space-like momenta. The
unparticle-like mode features a nearly flat potential with
\emph{spinodal instabilities} and a large expectation value. An
effective potential for the unparticle-like field is generated from
the Higgs potential, but with couplings suppressed by a large power
of the small see-saw ratio. The dispersion relation for the
Higgs-like mode features an imaginary part even at ``tree level'' as
a consequence of the fact that the unparticle field describes a
multiparticle continuum. Mixed unparticle-Higgs propagators reveal
the possibility of oscillations, albeit with short coherence
lengths. The results are generalized to the case in which the
unparticle features a mass gap, in which case a low energy MSW
resonance may occur for light-like momenta depending on the scales.
Unparticle-Higgs mixing leads to an \emph{effective unparticle} potential of the 
new inflation form. Slow roll variables are suppressed by see-saw ratios and 
the anomalous dimensions and favor a red spectrum of scalar perturbations consistent
with Cosmic Microwave Background (CMB) data. 
\end{abstract}

\pacs{12.60.-i; 14.80.-j; 98.80.Cq}

\maketitle

\section{Introduction}\label{sec:intro}
As evidence for physics beyond the Standard Model accumulates,
exploration of  extensions is becoming more focused by the
possibility of constraining them  with forthcoming collider
experiments. A recent proposal by Georgi  \cite{georgi} suggests that
a conformal sector with a non-trivial infrared fixed point coupled
to the standard model might be a possible extension with a wealth of
phenomenological consequences, some of which may be tested at the
Large Hadron Collider  \cite{cheung,raja,quiros}.  Early work by Banks
and Zaks  \cite{banks} provides a realization of a conformal sector
emerging from a renormalization flow towards the infrared below an
energy scale $\Lambda$ through dimensional transmutation, and
supersymmetric QCD may play a similar role  \cite{fox}.  Below this
scale there emerges an effective interpolating field, the unparticle
field, that features an anomalous scaling dimension  \cite{georgi}. 
 
  Recently various studies   recognized important
phenomenological  \cite{georgi,cheung,spin12} (for important  caveats see \cite{ira}),
astrophysical  \cite{raffelt,deshpande,freitas} and
cosmological  \cite{macdonald,davo,lewis,kame,he,kiku,xuelei}
consequences of unparticles, including Hawking radiation into unparticles \cite{dai} and
potentially relevant phenomenology in CP-violation \cite{zwicky1}, flavor physics \cite{flavor}
and low energy parity violation \cite{parity}.

A deconstruction program describes the unparticle from the coupling of a particle to a tower of a continuum of excitations \cite{micha}. Although this is an interesting interpretation of unparticles, the physics of anomalous dimensions arising from the exchange of
massless (conformal) excitations is an ubiquitous feature in \emph{critical phenomena},
a field that was already well developed before the advent of unparticle physics.
Anomalous dimensions in the fermion propagators in gauge theories 
had been understood in the mid 40's-50's \cite{bloch, bogo}, where multiple emissions and
absorptions of massless quanta leads  to anomalous dimensions. This is also well known
within the context of QCD \cite{Neubert:2007kh}. 

Critical phenomena associated with second order phase transitions provide a
natural realization of unparticle physics. Indeed, a scalar order parameter, such
as  the magnetization in a three dimensional Heisenberg ferromagnet, features anomalous
scaling dimensions at a critical point. That the multiple exchange of massless excitations leads to anomalous scaling exponents at critical points has been known since the 70's and well understood via renormalization group or large N resummations by the 80's \cite{amit, brezin}. Scale invariance appears as a
consequence of renormalization group flow towards an infrared fixed
point and the correlation functions of the order parameter scale
with anomalous scaling dimensions. This, of course was the original
motivation behind the Banks-Zaks suggestion within the context of
non-abelian gauge theories \cite{banks}. In critical phenomena,
integrating out degrees of freedom below a cutoff scale (in
condensed matter systems determined by the lattice spacing) down to
a renormalization scale $\Lambda$ yields an effective field that
describes long-wavelength phenomena below this scale. The effective
potential of the unparticle field is the Landau-Ginzburg free
energy, a functional of the order parameter, whose second derivative
with respect to the (scalar) order parameter (the unparticle field)
vanishes at the critical point; the effective potential
becomes \emph{flat}, reflecting the underlying scale invariance
emerging at the infrared fixed point \cite{amit,brezin}.

All of these predate the
``deconstruction'' interpretation, in some cases by decades. Although  the practitioners of ``unparticle-physics''
in the literature may prefer the ``deconstruction'' interpretation, we would like to emphasize that unparticle fields are ubiquitous in critical phenomena, and that the ``deconstruction'' interpretation is one, but by no means the only one. 
 
 
In this article, we are motivated to study unparticle physics by
the similarity between the effective field theory of single field
slow roll inflation as a paradigm for cosmological inflation whose
predictions are in remarkable agreement with WMAP
data \cite{WMAP3,WMAP5},  and a nearly critical theory. On the one hand, single field,
slow roll inflation is based on the dynamics of a scalar field, the
inflaton,whose evolution is determined by a fairly flat potential. The power
spectrum of inflaton fluctuations is nearly Gaussian and scale
invariant \cite{scott}, and non-linear couplings are small. In
ref. \cite{slowroll} it was argued that an effective field theory
description \emph{a l\'a} Landau-Ginzburg provides a compelling
description of single field slow roll inflation in which the
hierarchy of slow roll parameters emerges as a systematic expansion
in the number of e-folds and the non-linear couplings emerge as
see-saw like ratios of two widely different scales, the Hubble scale
during inflation and the Planck scale. The nearly Gaussian and scale
invariant spectrum of fluctuations, the flatness of the potential,
necessary to allow at least $60$ e-folds, with the concomitant
smallness of the mass of the inflaton field, and the smallness of the relevant couplings all suggest that perhaps a hidden scale (or conformal) invariant
sector is underlying the successful paradigm of slow roll inflation (see also  \cite{finnrob}).

In ref. \cite{slowroll}, this observation led to the suggestion that
perhaps, slow roll inflation is described by an effective field
theory \emph{near} a low energy fixed point, thus unparticle physics
may provide a framework for inflationary cosmology. Indeed,
recently in \cite{rich} some of the roles of unparticles in inflationary
cosmology were studied, along with the intriguing possibility that the unparticle field itself may be the inflaton.

On the other hand,   however, inflation \emph{cannot} be described
by an exactly scale invariant theory: the power spectrum of inflaton
fluctuations is not exactly scale invariant, and the inflaton
potential cannot be completely flat, since inflation must end,
eventually merging with standard Hot Big Bang cosmology, which in
turn implies
 that the inflaton mass cannot vanish, thus preventing the inflaton from being described by a conformally invariant unparticle
 field \cite{rich}.

 Thus a mechanism that would lead to a \emph{small} breaking of conformal
 invariance of the unparticle sector is sought. In ref. \cite{fox},
 it was recognized that if a scalar unparticle field couples to a
 Higgs field, a non-vanishing expectation value of the Higgs field
 leads to the explicit breakdown of the conformal symmetry in the
 unparticle sector. Although slow roll inflation is not exactly  scale
 invariant, it is nearly so, so that the explicit breaking of scale invariance
 must be such so as to lead to a nearly flat inflaton potential, leading in turn
to a nearly scale invariant spectrum of
 fluctuations \cite{scott}. This reasoning suggests that we should study the
 coupling between the unparticle and the Higgs field of the
 \emph{see-saw} form, just as in the case of neutrino mixing \cite{book1,book2,book3} (see  \cite{bosonicseesaw} for some work on this topic, and \cite{sannino} where the unparticle and Higgs particle
 are taken as composites. ).

 In this article, we focus on studying in detail the mixing between
 a scalar unparticle and a Higgs field in a spontaneously broken phase in flat space time as a
 prelude to dealing with inflationary cosmology. In particular we address
 the following issues.

\begin{itemize}
\item What are the consequences of mixing fields of \emph{different scaling
dimensions}? More specifically, using the language of neutrino
mixing, how do we compute the mixing angles and ``mass eigenstates''?
Although unparticle-Higgs mixing has been studied in the
literature \cite{fox,quiros}, to the best of our knowledge these
issues have not been addressed (although see the first reference in   \cite{bosonicseesaw}).

\item{ Consider a see-saw mass matrix between two \emph{canonical} scalar fields, one massless and one massive, namely
\be \Bigg(\begin{array}{cc}
      0 & m^2 \\
      m^2 & M^2
    \end{array} \Bigg) \,.\label{seeMM}\ee  For $M^2 \gg m^2$, it
    follows that there is one eigenvalue $\sim M^2$ corresponding to the massive scalar and
    another eigenvalue $\sim -m^4/M^2$ corresponding to the lighter
    state. However, this latter eigenvalue describes an
    \emph{instability}. Since a canonical scalar field is a special case of unparticle, it is natural to ask the questions if and how the fact
    that the unparticle field has non-canonical scaling dimension
    alters the see-saw mechanism and whether there is an instability
    as in the case of canonical fields.

    It is interesting to note that in the absence of underlying symmetries, the vanishing matrix element
    corresponding to the lighter scalar would be modified by
    radiative corrections. However, it is precisely the underlying
    conformal symmetry in the unparticle sector that guarantees the vanishing of that
matrix element.}
\end{itemize}

What we find in this study is quite interesting. First we see that the mixing between the unparticle and the Higgs field enjoys a number of  similarities with  the MSW phenomenon of neutrino mixing in a medium \cite{msw,book1,book2,book3}, namely the mixing angle \emph{depends on the energy}. This is a direct consequence of the non-canonical nature of the
 unparticle fields, with the hidden sector that lends the multiparticle nature to the unparticle interpolating field acting as a ``medium."
 We find the possibility of \emph{two MSW resonances} one at low and one at high energy.

We also show that  the combined unparticle-Higgs system exhibits
spinodal instability as well as a nearly flat potential. The
propagator of the diagonal field closest to the unparticle field
exhibits a \emph{pole} for space-like momenta (in Minkowski
space-time) which is exactly the signal of spinodal instability. For
small unparticle-Higgs mixing we show that this instability implies
that the field corresponding to the unparticle develops a large
expectation value and its potential is nearly flat.  This
instability is a remnant of the instability described by the see-saw
matrix (\ref{seeMM}). The unparticle-like field develops a potential
with self-interaction which is suppressed by a high power of the
see-saw ratio $m/M$.

Even when the unparticle-Higgs mixing is described by a \emph{linear coupling} between them, namely $\propto UH$, the propagator of the Higgs-like field features a \emph{complex} pole, the imaginary part (in Minkowski space-time) of which describes the \emph{decay} of the Higgs-like degree of freedom into unparticle-like degrees of freedom. The fact that this decay can happen even for \emph{linear coupling} reflects the fact that the unparticle field describes multi-particle states and the Higgs couples to a \emph{continuum} described by the spectral density of the unparticle.

We generalize the above results for the case when symmetry breaking in the Higgs sector induces a mass gap in the unparticle sector.

We find that unparticle-Higgs coupling leads to an \emph{unparticle effective potential} of the new
inflation form, with coefficients that are suppressed by the see-saw ratios and further suppression from
the anomalous dimensions for the resulting slow-roll parameters.

\section{Unparticle-Higgs Mixing}\label{sec:unhiggs}

The unparticle field describes a low energy conformal (or rather, a
scale invariant) sector \cite{georgi,banks,cheung}. The unparticle field scales with  an
 anomalous scale dimension that can be interpreted as a non-integral number of ``invisible'' particles \cite{georgi}.
This situation is akin to that of a scalar order parameter at a
non-trivial (Wilson-Fisher)
 infrared fixed point in critical phenomena \cite{amit,brezin}. At this critical point this sector becomes
scale invariant and correlation functions of the unparticle field
scale with anomalous dimensions. The anomalous scaling dimension
reflects the nature of the multiparticle intermediate states and the
unparticle propagator features a dispersive representation with a
spectral density that features anomalous scaling exponents and
describes branch cut singularities for time-like momenta.

We consider the following Euclidean-space Lagrangian for
unparticle-Higgs mixing \be L = \int d^4x \, d^4y \,
\frac{1}{2}\Big[ U(x)F(x-y)U(y)+
\Phi(x)(-\square)\delta^4(x-y)\Phi(y)\Big] + \int d^4x \, \Big[g
\Lambda \, U(x) \, \Phi^2(x) + V(\Phi)\Big], \label{UfiL}\ee where
$\Lambda$ has dimensions of mass (or momentum) and is a scale that
characterizes the unparticle-field \cite{georgi,banks,cheung}. The
unparticle field $U$ emerges as a composite interpolating field that
describes the infrared fixed point below this scale
\cite{georgi,banks,cheung},  and $g  \ll 1$ is a dimensionless
coupling.

The scalar potential $V(\Phi)$ features a symmetry breaking minimum
at $\varphi$, therefore we write \be \Phi = \varphi + H,
\label{higgs}\ee and in order to study the mixing between the
unparticle and the Higgs sector we keep only up to quadratic terms
in $U$ and $H$ in the above Lagrangian, \be L_{(2)} = \int d^4x \,
d^4y \, \frac{1}{2}\Big[ U(x)F(x-y)U(y)+ H(x)D(x-y)H(y)\Big] + \int
d^4x \Big[h \,U(x) +   m^2 \, U(x)H(x)   \Big], \label{UfiL2}\ee
where \be h = g  \Lambda \, \varphi^2 ~; \qquad m^2 =  2 g  \Lambda
\, \varphi \,.\label{para}\ee The vacuum expectation value of the
Higgs field breaks explicitly scale invariance in the unparticle
sector \cite{fox}.

We note that the Lagrangian density (\ref{UfiL}) can be extended to include a 
term of the form $ U^2(x) \Phi^2(x) $ which would result in an explicit mass term 
$M^2_U\, U^2(x)$ for the unparticle field when the Higgs acquires an expectation
value.  This explicit mass term, which obviously breaks conformal symmetry, moves
the unparticle threshold in the spectral density to $M^2_U$. Indeed the same massless
``hidden sector'' that gives the unparticle its multiparticle nature yields a
spectral density that now features a threshold at the value of the conformal breaking
mass $M^2_U$ \cite{fox}. We will consider this case explicitly in section (\ref{sec:mass}). 

The linear term in $U$ in (\ref{UfiL2}) leads to a tadpole
contribution to the unparticle field that requires
renormalization \cite{quiros}. In what follows we will neglect this
term and focus on the quadratic form to study the mixing.

The Euclidean space-time Fourier transforms of the non-local kernels
$F$ and $D$, denoted by $\mathcal{F}(p)$ and $\mathcal{D}(p)$, respectively,  are given by \cite{georgi,fox}
\bea \mathcal{F}(p) & = &  p^2  \,\Big(
\frac{p^2}{\Lambda^2}\Big)^{-\eta},
  \label{Fft}\\ \mathcal{D}(p) & = & p^2+M^2_H, \label{Dft}\eea
  where $ 0 \leq  \eta <  1 $\footnote{The anomalous dimension $\eta$ defined in this manner is \emph{twice} the critical
  exponent for the two point correlation function in a critical theory \cite{amit}.}.

  We will consider weak unparticle-Higgs mixing
and that the unparticle scale $\Lambda \gg
\varphi$ \cite{georgi,cheung}. These result in the following hierarchy of mass scales: \be m \ll M_H \ll \Lambda \,.\label{hie}\ee
  For $M_H \gg m$ the   mixing matrix will be of the see-saw form which is our primary interest for
  inflationary cosmology as discussed above. Furthermore,  consistent with the unparticle
  interpretation, the symmetry breaking scale of the Higgs sector must be below the unparticle scale so
  that the interpolating unparticle field is a suitable description of the hidden sector in the broken symmetry
  phase.

In absence of unparticle-Higgs interaction, the Euclidean propagator
for the unparticle field \be G_U(p) = \mathcal{F}^{-1}(p) = \frac{1}{p^2 \,\Big( \frac{p^2}{\Lambda^2}\Big)^{-\eta}}
\label{Uprop}\ee is normalized so that \be G_U(\Lambda) = \frac{1}{\Lambda^2} \label{norma}\ee as is usually done in critical
phenomena. This normalization fixes the wave function
renormalization constant at the scale $\Lambda$ \cite{amit}.
Alternatively the field can be normalized so that $d
G^{-1}_U(p^2)/dp^2\big|_{p^2=\Lambda^2}=1$ which differs from the
previous normalization by a finite wave-function renormalization
constant for $\eta <1$.  This field normalization differs from the
usually adopted one in the literature \cite{georgi}. We note that the
unparticle field $U$ normalized in this manner features
\emph{engineering} mass dimension $1$ just like an ordinary scalar
field, but \emph{scaling} mass dimension $1-\eta $ as befits a
conformal field with anomalous dimension $\eta$. Therefore the
\emph{engineering} mass-dimensions of $h$ and $m$ are $3$ and $1$,
respectively. It is convenient to pass on to Fourier transforms (in
Euclidean space-time) introducing the Fourier transforms of the
fields as $\widetilde{U},\widetilde{H}$ respectively in terms of
which the quadratic part of the action (\ref{UfiL2}) becomes \be S_{(2)}  = \int d^4p ~\frac{1}{2}~\Big(\wu(-p) \,\,\,
\wh(-p)\Big)~\Bigg(\begin{array}{cc}
                                                               \mf(p) & m^2 \\
                                                               m^2 &
                                                               \md(p)
                                                             \end{array}
                                                             \Bigg)~\Big(\begin{array}{c}
                                                                      \wu(p) \\
                                                                      \wh(p)
                                                                    \end{array}\Big).
\label{lagFT}\ee
The Lagrangian is
diagonalized in the basis of ``mass'' eigenstates (borrowing from
the language of neutrino mixing) $\Psi$ and $\chi$, related to the
unparticle and Higgs fields as \be \Bigg(
             \begin{array}{c}
               \wu(p) \\
               \wh(p) \\
             \end{array}
           \Bigg) ~ = \Bigg(
                        \begin{array}{cc}
                          C(p) & S(p) \\
                          -S(p) & C(p) \\
                        \end{array}
                      \Bigg)~\Bigg(
             \begin{array}{c}
               \Psi(p) \\
               \chi(p) \\
             \end{array}
           \Bigg), \label{trafo}\ee where \bea C(p)  & = &
           \frac{1}{\sqrt{2}}~\Bigg[1+ \frac{\md(p)-\mf(p)}{\big[\big(\md(p)-\mf(p) \big)^2 + 4 m^4
           \big]^{\frac{1}{2}}}
           \Bigg]^{\frac{1}{2}}\label{cofp}, \\ S(p)  & = &
           \frac{1}{\sqrt{2}}~\Bigg[1- \frac{\md(p)-\mf(p)}{\big[\big(\md(p)-\mf(p) \big)^2 + 4 m^4
           \big]^{\frac{1}{2}}}
           \Bigg]^{\frac{1}{2}} \label{sofp} \eea are
           effectively
           the cosine  ($C(p)$ )and sine ($S(p)$) of the ``mixing angle'' between the
           unparticle and Higgs fields.

           In terms of ``mass eigenstate''
             fields $\Psi,\chi$, we can write the action as
     \be S = \int d^4p ~ \frac{1}{2}~
           \Big(\Psi(-p)\,\,\,\chi(-p)\Big) ~\Bigg(
                                          \begin{array}{cc}
                                            G^{-1}_\Psi(p) & 0 \\
                                            0 & G^{-1}_\chi(p) \\
                                          \end{array}
                                        \Bigg) \Big(
                                                 \begin{array}{c}
                                                   \Psi(p) \\
                                                   \chi(p) \\
                                                 \end{array}
                                               \Big),
                                         \label{diagL}\ee where
\bea G^{-1}_\Psi(p) & = & \frac{1}{2} \Bigg( \mf(p)+\md(p)-
\Big[\big(\md(p)-\mf(p) \big)^2 + 4 m^4
           \Big]^{\frac{1}{2}}\Bigg),
\label{proppsi}\\G^{-1}_\chi(p) & = & \frac{1}{2} \Bigg(
\mf(p)+\md(p)+ \big[\big(\md(p)-\mf(p) \big)^2 + 4 m^4
           \big]^{\frac{1}{2}}\Bigg) \label{propchi}. \eea

           \section{MSW effect: resonances, mixing and oscillations}

           The similarity with the MSW effect of neutrinos in a
           medium \cite{msw} can be established   by defining a
           ``self-energy'' for the unparticle field \be \Sigma_U(p)
           = p^2 \Big[\Big( \frac{p^2}{\Lambda^2}\Big)^{-\eta} -1  \Big],
           \label{sigU}\ee which for $\eta \ll 1$ is reminiscent
           of the one-loop self energy of a scalar order parameter
           at an infrared  non-trivial critical point renormalized at a scale $\Lambda$, namely $\Sigma_U(p) \simeq
           -\eta p^2 \ln(p^2/\Lambda^2)$ \cite{amit}. The
           action (\ref{lagFT}) can   be written as
           \bea S =  \int
d^4p ~\frac{1}{2}~\Big(\wu(-p)\,\,\, \wh(-p)\Big)~\Bigg[\big(p^2+
  \frac{M^2_H}{2}\big)\,\mathbf{I} &+& \frac{1}{2} \Big[  M^4_H  +
  4m^4
\Big]^{\frac{1}{2}} \Bigg(
\begin{array}{cc}
                                                               -\cos(2\theta) & \sin(2\theta) \\
                                                               \sin(2\theta) &
                                                               \cos(2\theta)
                                                             \end{array}
                                                             \Bigg) \nonumber \\
                                                                    &+& \Bigg(
                                                                       \begin{array}{cc}
                                                                         \Sigma_U(p) & 0 \\
                                                                         0 & 0 \\
                                                                       \end{array}
                                                                     \Bigg)
                                                                     \Bigg]~\Big(\begin{array}{c}
                                                                      \wu(p) \\
                                                                      \wh(p)
                                                                    \end{array}\Big),
\label{lagFTmsw}\eea where $\mathbf{I}$ is the identity $2\times 2$
matrix and \bea \cos(2\theta) & = & \frac{M^2_H }{\Big[ M^4_H+ 4m^4
\Big]^{\frac{1}{2}}} \label{cos2},\\\sin(2\theta) & = &
\frac{2m^2}{\Big[  M^4_H+ 4m^4 \Big]^{\frac{1}{2}}}. \label{sin2}
\eea This action can be written in terms of a ``mixing angle in the
medium'' $\theta_m(p)$ as follows \bea L = \int d^4p
& &\frac{1}{2}~\Big(\wu(-p)\,\,\, \wh(-p)\Big) \times \nonumber \\
& &\Bigg[  \big(p^2+
\frac{M^2_H}{2}+\frac{\Sigma_U(p)}{2}\big)\,\mathbf{I} + \frac{1}{2}
\Big[( \Sigma_U(p)-M^2_H )^2+ 4m^4 \Big]^{\frac{1}{2}} \Bigg(
\begin{array}{cc}
                                                               -\cos(2\theta_m(p)) & \sin(2\theta_m(p)) \\
                                                               \sin(2\theta_m(p)) &
                                                               \cos(2\theta_m(p))
                                                             \end{array}
                                                             \Bigg)
                                                                     \Bigg] \nonumber\\
                                                                    & & \times  \Big(\begin{array}{c}
                                                                      \wu(p) \\
                                                                      \wh(p)
                                                                    \end{array}\Big),
\label{lagFTmsw2}\eea with \bea  \cos(2\theta_m(p)) & = & \frac{
M^2_H -\Sigma_U(p)  }{\Big[(\Sigma_U(p)- M^2_H  )^2+ 4m^4
\Big]^{\frac{1}{2}}}, \label{cos2m}\\\sin(2\theta_m(p)) & = &
\frac{2m^2}{\Big[( \Sigma_U(p)-M^2_H  )^2+ 4m^4
\Big]^{\frac{1}{2}}}. \label{sin2m} \eea This form makes manifest
that the multiparticle contribution that \emph{defines} the
unparticle can be associated with a ``medium effect'', namely the
degrees of freedom that have been integrated out leading to the
anomalous scaling dimension of the unparticle field, and leads to a
MSW phenomenon \cite{msw}, namely a
dependence of the mixing angle on the energy (here the Euclidean
momentum). From the expressions (\ref{cos2m},\ref{sin2m}) it is
clear that $C(p)$ and $S(p)$ of (\ref{cofp},\ref{sofp}) are
identical to $\cos(\theta_m(p))$ and $\sin(\theta_m(p))$,
respectively.

\begin{figure}
\begin{center}
\includegraphics[height=12cm,width=12cm,keepaspectratio=true]{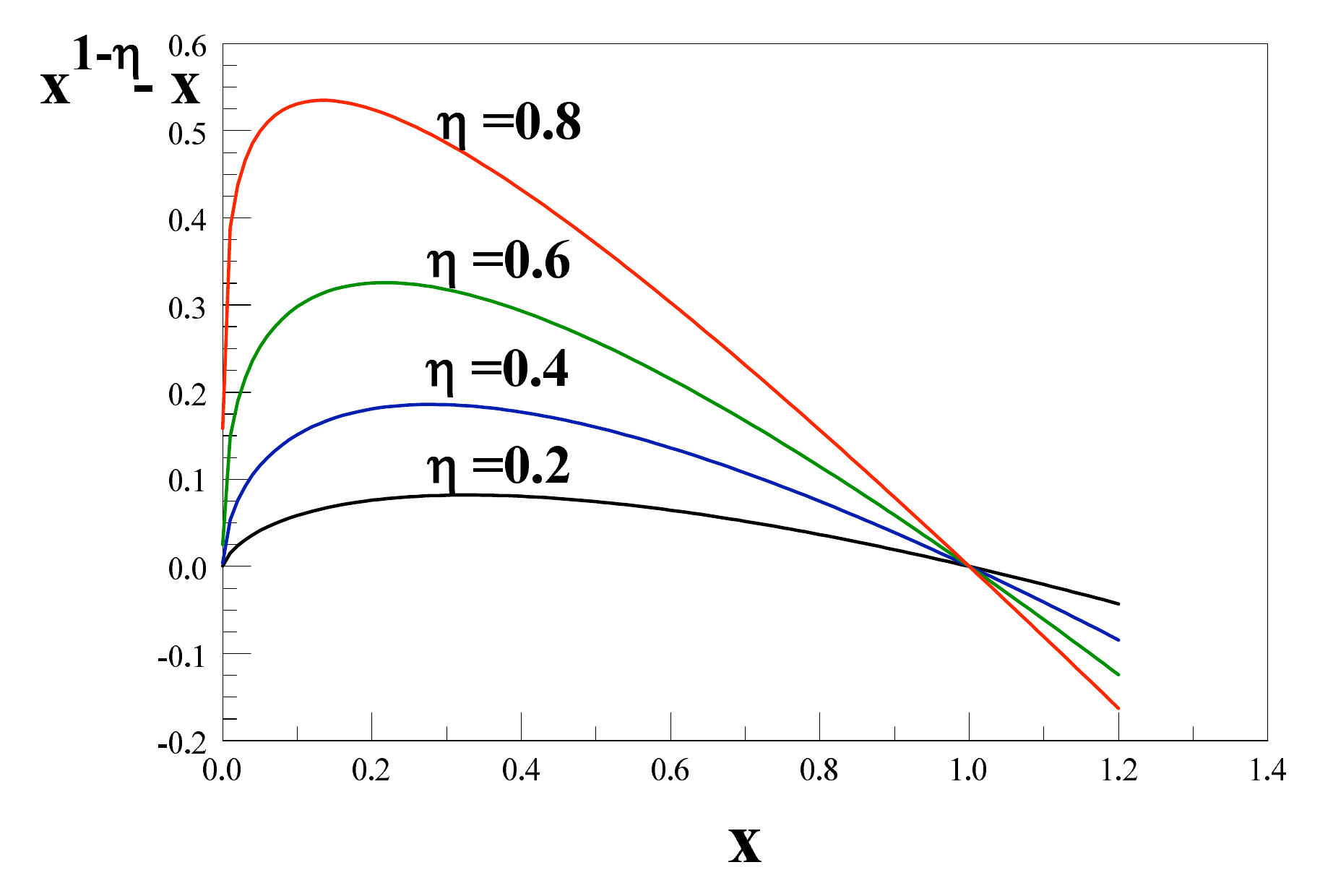}
\caption{The function $x (x^{-\eta}-1)$ vs $x$ for
$\eta=0.2,0.4,0.6,0.8$. } \label{fig:mswres}
\end{center}
\end{figure}

There is an MSW resonance phenomenon when the condition \be
\Sigma_U(p) = M^2_H  \label{mswres}\ee is fulfilled. It is
convenient to define \be x \equiv \frac{p^2}{\Lambda^2}
\label{defs}\ee in terms of which the resonance condition becomes
\be x \Big[x^{-\eta}-1\Big] = \frac{  M^2_H}{\Lambda^2}.
\label{condi}\ee The function $x (x^{-\eta}-1)$ is depicted in fig.
(\ref{fig:mswres}). From this figure it becomes clear that there are
\emph{two} MSW resonances whenever \be \frac{ M^2_H}{\Lambda^2} <
\eta~ (1-\eta)^{\frac{1-\eta}{\eta}}. \label{mswcon}\ee

Consider the case $\Lambda^2 \gg M^2_H  $ such that the condition
for MSW resonances (\ref{mswcon}) is fulfilled: there is a low
energy resonance ($p^2 \ll \Lambda^2$) at \be p^2 \simeq M^2_H
\Big[\frac{ M^2_H}{\Lambda^2}\Big]^\frac{\eta}{1-\eta}
\label{lowmsw}\ee and a high energy resonance ($p^2 \sim \Lambda^2$)
at \be p^2 \simeq \Lambda^2\Big[1-\frac{1}{2\eta}\frac{
M^2_H}{\Lambda^2}\Big]. \label{himsw}\ee Upon analytically continuing
from Euclidean to Minkowski momenta $p^2 \rightarrow
-(p^2_0-\vec{p}^{\,2}) -i0^+ $ we note that the low energy resonance
occurs near the light cone but for  slightly space-like momenta (in
the limit $  M^2_H \ll \Lambda^2$) whereas the high energy resonance
occurs at large space-like momenta.

\subsection{Mixing and oscillations}\label{mix} Unlike the case of neutrinos
wherein a single particle Fock representation is a suitable
description of the quantum mechanics of mixing and oscillations, the
fact that the unparticle field is an effective field that describes
multiparticle states prevents a similar analogy. However, we can
learn about mixing and oscillation phenomena by studying correlation
functions of the unparticle and Higgs fields. This is best achieved
by introducing sources $J_{U,\Phi}$ conjugate to the respective
fields and a generating functional $Z[J_U,J_\Phi]$ that yields the
correlation functions through functional derivatives.  It is
straightforward to obtain the generating functional by inverting the
quadratic form in the action (\ref{lagFT}), we find \be \ln
Z[J_U,J_\Phi] =  \frac{1}{2} \int d^4 p \Bigg(J_U(-p)\,\,\,J_{\Phi}(-p)
\Bigg) \left(G_\Psi(p)\,G_\chi(p) \Bigg( \begin{array}{cc}
                                                                \mathcal{D}(p) & -m^2 \\
                                                                -m^2 & \mathcal{F}(p)
                                                              \end{array}
 \Bigg)\right) \Bigg( \begin{array}{c}
                 J_U(p) \\
                 J_\Phi(p)
               \end{array}
 \Bigg), \label{ZJJ}\ee where the
 Green's functions of ``mass eigenstates'' $G_{\Psi,\chi}(p)$ are given by eqns. (\ref{proppsi},\ref{propchi}). For convenience of notation, we introduce
 \bea \alpha(p)  & = &  \frac{1}{2} \Bigg( \mathcal{F}(p) + \mathcal{D}(p)\Bigg) = p^2 + \frac{M^2_H}{2} + \frac{\Sigma_U(p)}{2} ,\label{alfa}\\ \beta(p) & = & \frac{1}{2} \Big[\big(\mf(p)-\md(p) \big)^2 + 4 m^4
           \Big]^{\frac{1}{2}}=  \frac{1}{2} \Big[(\Sigma_U(p)-
M^2_H )^2+ 4m^4 \Big]^{\frac{1}{2}}, \label{beta} \eea in terms of
which $G^{-1}_{\Psi}(p) = \alpha(p)+\beta(p),\ G^{-1}_{\chi}(p) =
\alpha(p)-\beta(p)$. The diagonal and off-diagonal correlation
functions in the unparticle-Higgs basis are given by \bea \langle
U(p) U(-p)\rangle &
= & \frac{\mathcal{D}(p)}{\alpha^2(p)-\beta^2(p)}, \label{Ucorre}\\
\langle \Phi(p) \Phi(-p)\rangle & = &
\frac{\mathcal{F}(p)}{\alpha^2(p)-\beta^2(p)}, \label{Ficorre} \\
\langle U(p) \Phi(-p)\rangle & = & -
\frac{\beta(p)~\sin(2\theta_m(p))}{\alpha^2(p)-\beta^2(p)} = \frac{1}{2}\sin(2\theta_m(p))\Bigg[\frac{1}{G_{\chi}(p)} -
\frac{1}{G_{\Psi}(p)}\Bigg]. \label{offcorre}\eea

The off-diagonal propagator (\ref{offcorre}) is exactly of the same form of the mixed propagators
for two neutrinos in the flavor basis \cite{boyaneutrino} where $G^{-1}_{\Psi,\chi}$ correspond to the
propagators of mass eigenstates. In the case of neutrino mixing, the transition probability emerges
directly from the off-diagonal correlation function as follows. Analytically continuing to Minkowski space-time,
each propagator has a simple pole at the values of $p_0 = E_1(\vp),E_2(\vp)$ respectively. Then the
time evolution of the off-diagonal correlator is obtained by performing an inverse Fourier transform in time, which in turn yields
 the time dependence for the off-diagonal correlator $$\propto \sin(2\theta_m(p)) \Bigg[e^{i E_1(\vp)t}-e^{iE_2(p)t}\Bigg]\,.$$
  The transition probability is obtained from the absolute value squared of the off-diagonal correlator \cite{boyaneutrino}
   $\mathcal{P}(t) \propto  \sin^2(2\theta_m(p)) \sin^2 \big[(E_1(\vp)-E_2(\vp))t/2\big]$.

 Hence the off-diagonal correlation function (\ref{offcorre}) describes  the generalization of mixing and
 oscillations for unparticle-Higgs mixing. For weak unparticle-Higgs coupling, we expect the propagators
  to feature singularities near those   corresponding to the unparticle cut beginning at $p^2 = 0$ and Higgs pole at $p^2=M^2_H$ respectively
 in Minkowski space-time. Therefore, unlike the case of almost degenerate neutrinos, the unparticle-Higgs oscillation will not be coherent over long
 space-time intervals. Instead, dephasing occurs on space-time scales of the
 order of the Compton wavelength of the Higgs-like mode $\sim 1/M_H$ and is
 further suppressed by the decay of this mode (see below).

           \section{Singularity structure: poles and cuts} The
           singularity structure of the propagators for the ``mass
           eigenstates'' is obtained from the conditions \be
           G^{-1}_\Psi(p) =0,\  G^{-1}_\chi(p) =0 \,.\label{poles}
           \ee Both conditions can be combined into \be
           \mathcal{F}(p)\,\mathcal{D}(p) = m^4. \label{poleconds}\ee
           For $m^2 \ll M^2_H \ll \Lambda^2$ we expect to find
           singularities near the Higgs ``mass shell'' $p^2 \sim -
           M^2_H$ and near the beginning of the massless threshold
           for the unparticle $p^2 \sim 0$.

           \vspace{2mm}
\begin{itemize}

           \item Higgs-like pole: To find the position of the
           singularity near the Higgs mass shell we write \be
           p^2+M^2_H \equiv M^2_H ~\Delta ~~;~~ \Delta \ll 1
           \label{higgspole}\ee The condition (\ref{poleconds})
           yields \be \Delta \simeq - \frac{m^4}{M^4 _H} ~\Bigg(
           \frac{-M^2_H}{\Lambda^2}\Bigg)^\eta + \mathcal{O}\Bigg(\frac{m^8}{M^8_H}\left(\frac{- M_H^4}{\Lambda^4}\right)^{\eta} \Bigg)\,.\label{delta}\ee
           Upon analytic continuation to Minkowski
           space time $p^2_E \rightarrow -p^2_M; M^2_H \rightarrow M^2_H-i 0^+$ the pole becomes complex with
           an imaginary part \be \Gamma_H \simeq   \frac{m^4}{2\,M^3 _H} ~\Bigg(
           \frac{M^2_H}{\Lambda^2}\Bigg)^\eta~\sin(\pi \eta)\,. \label{width}\ee

           This is a complex pole of $G_\chi(p)$ given by (\ref{proppsi}). From the expressions
           (\ref{cofp},\ref{sofp}) with $p^2 \sim -M^2_H,\ \mathcal{D}(p) \sim 0,$ and  $\mathcal{F}^2(p)\gg m^4$, it follows that
           $C(p) \sim 1,\ S(p) \sim 0$ and the
           complex pole in $G_\chi(p)$ describes a Higgs-like \emph{unstable} particle. What is remarkable in this result is the fact that
           the unparticle-Higgs coupling is \emph{linear} in both fields, and the decay width emerges at ``tree
           level''. This is in marked contrast with the decay
           rate from non-linear couplings studied in
           ref. \cite{quiros}. If the unparticle field were canonical, such coupling would
           not result in an imaginary part in the propagator of the Higgs-like mode at tree level.
            However, the unparticle field is an interpolating field
           that describes a multiparticle composite with a continuum spectral density. Since the threshold begins at $p^2=0$, it follows
           that upon coupling the fields,
           the Higgs-pole on the positive real axis in the $p^2$ plane (in Minkowski space time)
           is actually embedded in the continuum of states described by the unparticle field, resulting in the motion of this pole off the
           physical sheet into a second or higher Riemann sheet.  The imaginary part just describes the decay of the Higgs at
           ``tree'' level.

           We note that the real part of the pole receives a finite
           mass renormalization, and the real part is light-like, therefore far
           away from the MSW resonance region which occurs for
           space-like momenta.

           \item Near unparticle threshold: The singularity near the unparticle threshold at $p^2 =0$ can be found by setting $p^2 =0$ in
           $\mathcal{D}(p)$ and corresponds to a singularity in
           $G_{\Psi}(p)$ given by eqn. (\ref{proppsi}).

           Defining $x = p^2/\Lambda^2 \ll 1$  the condition (\ref{poleconds}) yields  \be x^{1-\eta}
           \simeq  \frac{m^4}{M^2_H \Lambda^2}  \Rightarrow p^2 \simeq \frac{m^4}{M^2_H}~\Bigg(\frac{m^4}{M^2_H \Lambda^2}\Bigg)^{ \eta/(1-\eta)}
            \label{unpole}\ee For $\eta=0$ one recovers the negative eigenvalue $-m^4/M^2_H$ of the  see-saw mass matrix (\ref{seeMM}).
            Furthermore it is clear that although this pole is space like, it is well below the
            position of the low energy MSW resonance (\ref{lowmsw}).

            This
           pole on the positive real axis in Euclidean $p^2$
           describes an \emph{instability} since upon analytic continuation back to Minkowski space-time
           $p^2 \rightarrow -p^2= -(\omega^2-k^2)$ corresponding to
           frequencies \be \omega(k) = \sqrt{k^2 -\mathcal{M}^2} ~~;~~
           \mathcal{M}^2 = \frac{m^4}{M^2_H}~\Bigg(\frac{m^4}{M^2_H \Lambda^2}\Bigg)^{ \eta/(1-\eta)}\,. \label{spino}
           \ee These become imaginary in the  \emph{band of spinodally unstable
           modes} \cite{spino} \be 0 < k^2 \leq \mathcal{M}^2 \,.
           \label{banda}\ee corresponding to spinodal
           instabilities \cite{spino}.

           \end{itemize}

            Thus we see that the instability obtained from the see-saw mass matrix (\ref{seeMM}) is reproduced but with
            a coefficient that depends on the ratio of scales and the anomalous
            dimension. For this pole the mixing angles $C(p) \sim
            1,\ S(p) \sim 0$ and the $\Psi$ field  is identified with an
            unparticle-like mode, the pole for $p^2 <0 $ in
            Minkowski is on the opposite side of the branch cut
            singularity for $p^2 >0$ and is isolated from the
            unparticle continuum.

            The spinodal instabilities signal that the ``potential''
            associated with the $\Psi$ (unparticle-like) mode
            features a minimum away from the origin and the
            instabilities reflect the ``rolling'' of the expectation
            value of the $\Psi$ field towards the
            minimum \cite{rolling}.

            The unparticle-Higgs mixing leads to a potential for the
            unparticle-like field $\Psi$. Consider the Higgs
            self-interaction $\lambda H^4$
            from (\ref{trafo}) \be \lambda H^4 = \lambda
            \Bigg(C(p)\chi(p)-S(p)\Psi(p)\Bigg)^4 \simeq
            \lambda S^4(p)\Psi^4(p), \label{psipot} \ee where we have focused
            on the direction along which $\chi =0$. For the unstable pole we can  set $p \simeq 0$
            in the expression for the mixing angle (\ref{sofp} or \ref{sin2m}) and
            we find \be S(p) \sim \theta_m(p\sim 0) \simeq
            \frac{m^2}{M^2_H}\;, \label{sofp0} \ee combining this result with eqn. (\ref{spino}) leads to the \emph{effective unparticle} 
            potential for the $\Psi$ mode \be \mathcal{V}(\Psi)
            \simeq \mathcal{V}(0) -\frac{\mathcal{M}^2}{2} \, \Psi^2+\lambda \frac{m^8}{M^8_H}\, \Psi^4 \,. \label{psipot}\ee which is of the form describing new inflation. 
            Thus we see that the effective self-coupling for the
            $\Psi$ mode (unparticle-like) is a see-saw ratio of
            scales, which for $m\ll M_H$ consistent with small
            breaking of conformal invariance, entails that
            $\mathcal{V}(\Psi)$ is very shallow. Even for $\lambda \sim 1$, the quartic
            self-coupling for the unparticle field is a large power
            of the see-saw ratio $m/M_H$ very similarly to the
            effective field theory approach to slow roll inflation
            discussed in ref. \cite{slowroll}.

            The expectation
            value of $\Psi$ is obtained by balancing the quadratic
            term whose coefficient is determined by the unstable
            pole (\ref{unpole}) and the quartic term (\ref{psipot}); we
            find \be \langle \Psi \rangle \simeq
            \frac{M^3_H}{\lambda\,m^2} \Bigg(\frac{m^2}{M_H \Lambda}\Bigg)^{ \eta/(1-\eta)}\,. \label{vevPsi}\ee

            It is interesting to note that the spinodal instabilities exist independently of the specific form of the Higgs potential.
             Even if we replace the Higgs and its $\lambda H^4$ self-interaction with a scalar field with no potential,
             these instabilities persist. In this latter case, the unparticle field will acquire a potential by the virtue of the
              Coleman-Weinberg mechanism \cite{Coleman:1973jx} and the instabilities reflect the rolling toward the minimum of this potential.

              \medskip

            \section{Unparticle mass gap}\label{sec:mass}
            In the study presented in the previous section, we have
            neglected a mass gap for the unparticle field. 
            
            As mentioned in section (\ref{sec:unhiggs}), a term $U^2 \Phi^2$ in the
            unparticle-Higgs Lagrangian yields an explicit mass term for the unparticle field, which
            manifestly breaks conformal invariance. The emission and absorption of massless excitations
            of the ``hidden'' conformal sector that leads to the multiparticle nature of unparticles, 
            results in that this mass term moves the threshold away from $p^2=0$ to $p^2 = M^2_U$ in
            the complex $p^2$ plane.  
            
            In
            ref. \cite{fox}, a simple manner to introduce a mass scale
            to break conformal invariance in the unparticle sector
            was introduced by modifying the spectral representation
            of the unparticle propagator. Such a modification was also
            used in the study in ref. \cite{quiros} where it is
            argued that the mass gap in the spectral representation
            arises   from unparticle-Higgs coupling. The
            introduction of an unparticle mass gap $M_U$ results in a
            spectral density that features a branch cut beginning at
            a threshold $p^2 = -M^2_U$ in Euclidean momentum. The
            spectral density featuring a branch discontinuity
            with threshold $M^2_U$ resulting in anomalous scaling
            dimensions is reminiscent of the
            Bloch-Nordsieck \cite{bloch}
            resummation of the emission and absorption of nearly collinear
            (soft) photons in quantum electrodynamics \cite{bogo} where the fermion
            propagator acquires an anomalous dimension from infrared threshold divergences \cite{bogo}.

             In
            ref. \cite{infradiv} it was shown that the
            renormalization group resummation of the infrared divergences
            arising from the emission and absorption of soft
            massless quanta by a massive field is equivalent to the
            Bloch-Nordsieck resummation and yields precisely a
            spectral density with a branch cut beginning at a
            threshold given by the mass of the particle that emits
            and absorbs the soft quanta.

            Motivated by these cases and following refs.\cite{fox,quiros}, we include an unparticle mass gap $M_U$ by modifying the
            $\mathcal{F}(p)$ in eqn. (\ref{Fft}) to \be
            \mathcal{F}(p) = (p^2+M^2_U)\Bigg[
            \frac{p^2+M^2_U}{\Lambda^2} \Bigg]^{-\eta}. \label{Ugap} \ee
This is the type of inverse propagator obtained by a renormalization
group or alternatively a Bloch-Nordsieck resummation of infrared
divergences arising from emission and absorption of soft massless
quanta \cite{infradiv}. Consistent with a small breaking of conformal
invariance, we focus on the case in which $M_U \ll M_H \ll \Lambda$.

The analysis presented in the previous sections can be followed by
replacing the unparticle self-energy (\ref{sigU}) by \be \Sigma_U(p)
           = (p^2+M^2_U) \Bigg[\Bigg( \frac{p^2+M^2_U}{\Lambda^2}\Bigg)^{-\eta} -1  \Bigg], \ee
and the action (\ref{lagFTmsw2}) by
 \bea L = & & \int d^4p \,\,\,
\frac{1}{2}~\Big(\wu(-p)\,\,\, \wh(-p)\Big) \times \nonumber \\
& & \Bigg[  \big(p^2+
\frac{M^2_H+M^2_U}{2}+\frac{\Sigma_U(p)}{2}\big) \mathbf{I}  \, +
\frac{1}{2} \Big[( \Sigma_U(p)+M^2_U-M^2_H )^2+ 4m^4
\Big]^{\frac{1}{2}}  \Bigg(
\begin{array}{cc}
                                                               -\cos(2\theta_m(p)) & \sin(2\theta_m(p)) \\
                                                               \sin(2\theta_m(p)) &
                                                               \cos(2\theta_m(p))
                                                             \end{array}
                                                             \Bigg)
                                                                     \Bigg] \nonumber\\
                                                                     & & \times \, \Big(\begin{array}{c}
                                                                      \wu(p) \\
                                                                      \wh(p)
                                                                    \end{array}\Big),
\label{lagFTmsw22}\eea with the ``in medium'' mixing angles
determined by \bea \cos(2\theta_m(p)) & = & \frac{ M^2_H-M^2_U
-\Sigma_U(p) }{\Big[(\Sigma_U(p)+M^2_U- M^2_H )^2+ 4m^4
\Big]^{\frac{1}{2}}}, \label{cos2m2}\\\sin(2\theta_m(p)) & = &
\frac{2m^2}{\Big[( \Sigma_U(p)+M^2_U-M^2_H  )^2+ 4m^4
\Big]^{\frac{1}{2}}}. \label{sin2m2} \eea

The condition for an MSW resonance now becomes \be \Sigma_U(p) = M^2_H-M^2_U \equiv \delta M^2  \,.\label{resgap}\ee Upon introducing
the variable $x = (p^2+M^2_U)/\Lambda^2$ this condition becomes  \be
x \Big[x^{-\eta}-1\Big] = \frac{  \delta M^2}{\Lambda^2},
\label{condi2} \ee which is again satisfied with \emph{two}
resonances for
\be 0 < \frac{ \delta M^2 }{\Lambda^2} < \eta~
(1-\eta)^{\frac{1-\eta}{\eta}}. \label{mswcon2}\ee
For $\Lambda \gg
M^2_H \gg M^2_U$ the resonances occur for  \be p^2 \simeq -M^2_U +
\delta M^2  \Big[\frac{ \delta M^2
}{\Lambda^2}\Big]^\frac{\eta}{1-\eta}, \label{lowmsw2}\ee
and
 \be
p^2 \simeq -M^2_U+\Lambda^2\Big[1-\frac{1}{2\eta}\frac{ \delta M^2
}{\Lambda^2}\Big]. \label{himsw2}\ee Upon analytically continuing
from Euclidean to Minkowski momenta $p^2 \rightarrow
-(p^2_0-\vec{p}^{\,2}) -i0^+ $ we note that whereas the high energy
resonance (\ref{himsw2}) occurs for space-like momenta there is the
tantalizing possibility that the low energy resonance
(\ref{lowmsw2}) \emph{could} occur  at light-like momenta. Obviously
whether or not this possibility is realized depends on the details
of the scales.

The results for mixing and oscillations obtained in section
(\ref{mix}) remain the same with obvious modifications in the mixing
angles and propagators.

\subsection{Singularity structure} The singularities in the
propagators lead to the same condition (\ref{poleconds}) as for
$M^2_U=0$.

\begin{itemize}
\item Higgs-like pole:  we write again \be
           p^2+M^2_H \equiv M^2_H ~\Delta ~~;~~ \Delta \ll 1
           \label{higgspole2}\ee and the  condition
           (\ref{poleconds}) now
           yields \be \Delta \simeq - \frac{m^4}{M^2_H \,\delta M^2} ~\Bigg(
           \frac{-\delta M^2}{\Lambda^2}\Bigg)^\eta.\label{delta2}\ee

           Upon analytic continuation to Minkowski
           space time $p^2_E \rightarrow -p^2_M; M^2_H \rightarrow M^2_H-i 0^+$ the pole becomes complex with
           an imaginary part \be \Gamma_H \simeq   \frac{m^4}{2\,M_H \, \delta M^2} ~\Bigg(
           \frac{\delta M^2}{\Lambda^2}\Bigg)^\eta~\sin(\pi \eta)\,\Theta(M^2_H -M^2_U) \,.
           \label{width2}\ee The $\Theta(M^2_H-M^2_U)$ in
           (\ref{width2}) results from the fact that now the continuum
           spectral weight for the unparticle has a threshold at
           $M^2_U$. Thus, if $M_H < M_U$ the pole in the bare
           Higgs propagator along the real axis in the Minkowski
           $p^2$ plane, lies \emph{below} the unparticle continuum
           and the Higgs-like particle is stable. For $M_H > M_U$
           this (bare) pole is in the unparticle continuum and moves
           off into an unphysical sheet indicating a decay width for the Higgs-like particle.

           \item Near unparticle threshold:   Defining   $x = (p^2+M^2_U)/\Lambda^2 \ll 1$  the condition
           (\ref{poleconds}) yields  \be x^{1-\eta}
           =  \frac{m^4}{\delta M^2  \Lambda^2}  \Rightarrow p^2
           \simeq \frac{m^4}{\delta M^2}~\Bigg(\frac{m^4}{\delta M^2 \Lambda^2}\Bigg)^{
           \eta/(1-\eta)}\,.
            \label{unpole2}\ee For $M^2_H > M^2_U$ this pole along
            the real axis in Euclidean space, again indicates
            spinodal instabilities. The band of spinodally unstable
            wave vectors is the same as in eqn. (\ref{banda}) but with
            \be \mathcal{M}^2 = \frac{m^4}{\delta M^2}~\Bigg(\frac{m^4}{\delta M^2 \Lambda^2}\Bigg)^{
           \eta/(1-\eta)} \,.\label{spinoband2}\ee If $M^2_U >
           M^2_H$ the value of $p^2$ becomes complex, indicating the
           possibility of the decay of the unparticle-mode. While
           this possibility is interesting and deserves to be
           explored in its own right, we are primarily interested in
           the case $M^2_H >> M^2_U$ because   this case corresponds
           to the  see-saw mechanism with two widely different mass
           scales that may be of relevance for inflationary
           cosmology.
\end{itemize}

           Therefore we conclude that in the relevant case $M^2_H > M^2_U$ the  results
           obtained are a straighforward generalization of the case
           $M^2_U=0$ with the same features, namely, a band of
           spinodally unstable wavevectors, a shallow effective
           potential for the unparticle mode and a width for the  Higgs-like particle
           indicating its decay into unparticle modes as a
           consequence of the multiparticle continuum described by
           them.
           
           Furthermore the effective unparticle potential is of the same form as (\ref{psipot}) but
           with the replacement $M^2_H \rightarrow \delta M^2$ with similar conclusions. 
                 
     \section{ Potential consequences for slow roll inflation} 
     Although the WMAP data \cite{WMAP3,WMAP5} rules out a \emph{purely quartic} inflaton potential, a systematic analysis combined with
              Markov-chain Montecarlo of the available data from the CMB and large
              scale structure shows that a new-inflation type potential with a \emph{non-vanishing} mass
              term, just as that given by eqn. (\ref{psipot}) fits the data remarkably well \cite{newinfla}. Therefore identifying the unparticle with the inflaton field \cite{rich}, with the effective inflaton potential given by (\ref{psipot}), we can advance some important preliminary consequences for slow roll inflation. Both lowest order slow roll parameters \be  \epsilon_v   =  \frac{M^2_{Pl}}{2}~\Bigg(\frac{\mathcal{V}'}{\mathcal{V}}\Bigg)^2  ~~;~~
               \eta_v   =   M^2_{Pl}~ \frac{\mathcal{V}''}{\mathcal{V}} \label{slowroll} \ee involve
               $\mathcal{M}^2$ for slow roll initial conditions ($\Psi \sim 0$). We see from (\ref{spino})
               that the slow roll parameters are determined $\mathcal{M}^2$ (for a general dependence
               see \cite{newinfla}).  This term is suppressed by the small see-saw ratio $m^2/M^2_H$, but
               also  for a non-trivial  unparticle anomalous dimension $0 < \eta < 1$
               it is \emph{further} suppressed by the factor \[ \Bigg(\frac{m^2}{M_H \Lambda}\Bigg)^{ \eta/(1-\eta)},\] making the slow roll parameters \emph{smaller}. To lowest order in slow
               roll parameters, the scalar index of the
               power spectrum is $n_s = 1+2\eta_v-6\epsilon_v$ \cite{WMAP3,WMAP5}. We note that whereas
               $\epsilon_v$ is manifestly positive, the sign of $\eta_v$ determines whether the power spectrum is red or blue tilted. The effective unparticle potential (\ref{psipot}) distinctly yields a red tilt which is consistent with the results from WMAP \cite{WMAP3,WMAP5} .


               We would like to emphasize that here, we have used slow roll parameters for a field with canonical kinetic term. This is done to give an indication that simply based on the potential, an unparticle field description of the inflaton can yield a nearly scale invariant power spectrum. However, as the kinetic term of the unparticle-like field is not canonical, a deeper understanding of the slow roll conditions in this case is required. This will be explored in a future work. It is conceivable that the next generation of CMB observations may yield information on the unparticle anomalous exponent.          
           
           
\section{Conclusions}

 Motivated by the successful and predictive paradigm of slow roll inflation,
 we explored the possibility that unparticle physics may yield to an
 underlying understanding of the main features of slow roll
 inflation: a fairly flat potential, nearly Gaussian and scale
 invariant spectrum of fluctuations. We identify these as hallmarks
 of a nearly critical theory. Thus, unparticle physics, describing a
 a conformally invariant theory appears as a natural candidate for
 an underlying theory of slow roll inflation. However, conformal
 invariance must be explicitly broken but in a ``small'' manner,
 since inflation must end. Coupling unparticle and Higgs fields in a
 broken symmetry phase of the Higgs sector leads to an explicit
 breaking of conformal symmetry in the unparticle sector \cite{fox}.
 However, we seek a mechanism that yields \emph{small} breaking of
 this symmetry in the form of a weak coupling between unparticle and
 Higgs sectors and a see-saw type mixing matrix between them. Thus
 we study a model of a scalar unparticle weakly coupled to a Higgs
 field in a broken symmetry phase. The expectation value of the
 Higgs leads to  linear Higgs-unparticle coupling and a see-saw type
 mixing matrix. We find a wealth of phenomena possibly relevant to
 inflationary cosmology but also of intrinsic interest.

 As a consequence of  the unparticle field being defined by non-canonical
 scaling dimensions we find that the mixing angles depend on the
 momentum four vector leading to an MSW effect: the hidden sector
 that is integrated out to define the interpolating unparticle field
 acts as a medium. We find \emph{two} MSW resonances, one at low and one
 at high energy respectively. For low momentum we find isolated
 poles in the unparticle-like mode away from the branch cuts that
 characterize the unparticle spectrum. These poles describe
 \emph{spinodal instabilities} and indicate that the unparticle
 field acquires an expectation value. Indeed, the Higgs potential
 generates an effective potential for the unparticle-like field
 because of the mixing. We find that even for a strongly self-coupled Higgs
 sector, the self-couplings of the unparticle-like mode are
 suppressed by large powers of the see saw ratios. The instability
 and small self-couplings both entail a nearly flat potential for
 the unparticle-like field, a hallmark of slow roll inflation.
 We also find the remarkable result that the propagator for the Higgs-like mode
 features a complex pole, whose imaginary part determines the decay
 rate. We emphasize that this is a ``tree-level'' effect, \emph{not a
 result of non-linear couplings} and is a consequence of the
 continuum in the spectral representation of the unparticle field.
 The pole in the bare Higgs propagator becomes embedded in the
 continuum of the unparticle field upon their coupling, even for a
 \emph{linear coupling}.

 Unparticle-Higgs mixing also leads to oscillation phenomena just as
 in the case of neutrino mixing. However, because of the large
 difference in scales, oscillations decohere on short space-time
 scales of the order of the Compton wavelength of the Higgs
 particle.

 The results obtained for a massless scale invariant unparticle
 field were then generalized to the case in which there is an unparticle
 mass gap $M_U$. A see-saw mechanism consistent with a slow roll
 picture requires that the unparticle mass gap be much smaller than
 the Higgs mass, in which case all of the results of the massless
 unparticle case translate with minor modifications to the case of
 a massive threshold for the unparticles.

 A major result of this exploration is that an unparticle field
 weakly coupled to a Higgs particle yields a remarkable similarity
 to the main features of slow roll inflation and could possibly
 provide an underlying justification for the slow roll paradigm. Unparticle-Higgs see-saw type coupling yields an \emph{effective  unparticle potential} of
 the new inflation form with coefficients that are suppressed by see-saw ratios. Simply based on this potential, the \emph{further} suppression by the anomalous dimension of the unparticle
 field might lead to smaller departures from scale invariance and to a red spectrum consistent with the WMAP results on the scalar spectral index. 
 
 Since the unparticle field features non-canonical
 kinetic terms, we envisage these models as possible alternatives to
 Dirac-Born-Infeld (DBI) inflationary proposals \cite{dbi}. An important follow-up will be to understand the influence of the non-canonical kinetic
 terms for the unparticle-like field so as to establish a firmer correspondence with
 the dynamics of slow roll inflation. Non-canonical kinetic terms in other contexts have been studied in
 refs. \cite{nonstandardkinetic} and we will explore these aspects along with loop corrections in future work.
  
 We focused our study in Minkowski space time as a prelude to a more
 comprehensive exploration of inflationary cosmology. The results
 obtained here are
 most certainly encouraging and suggest that further exploration
 within the inflationary context is
 worthwhile.

\begin{acknowledgements}
D.B. acknowledges support from the U.S. National Science Foundation through grant No:
 PHY-0553418. R. H. and J. H. acknowledge support from the DOE through grant DE-FG03-91-ER40682.
 J. H. is also supported by the De Benedetti Family Fellowship for Physics.
 \end{acknowledgements}

\end{document}